\documentclass[aps,prl,twocolumn,superscriptaddress,notitlepage,nofootinbib]{revtex4-1}
\usepackage{amsmath}
\usepackage{bm}
\usepackage{graphicx}
\usepackage{color,soul}
\usepackage{float}
\usepackage{multirow}
\usepackage[inline]{enumitem}
\usepackage{cancel}
\usepackage[normalem]{ulem}

\def\bea#1\eea{\begin{align}#1\end{align}}

\newcommand{\bef}{\begin{figure}[h!tb]\centering}
\newcommand{\eef}{\end{figure}}

\allowdisplaybreaks

\begin{document}
\title{Jet Charge: A Flavor Prism for Spin Asymmetries at the EIC}
\author{Zhong-Bo Kang}
\email{zkang@physics.ucla.edu}
\affiliation{Department of Physics and Astronomy, University of California, Los Angeles, California 90095, USA}
\affiliation{Mani L. Bhaumik Institute for Theoretical Physics,
University of California, Los Angeles, California 90095, USA}
\affiliation{Center for Frontiers in Nuclear Science, Stony Brook University, Stony Brook, New York 11794, USA}
\author{Xiaohui Liu}
\email{xiliu@bnu.edu.cn}
\affiliation{Center of Advanced Quantum Studies, Department of Physics,
Beijing Normal University, Beijing 100875, China}
\affiliation{Center for High Energy Physics, Peking University, Beijing 100871, China}
\author{Sonny Mantry}
\email{mantry147@gmail.com}
\affiliation{Department of Physics and Astronomy, 
                   University of North Georgia,
                   Dahlonega, Georgia 30597, USA}
\author{Ding Yu Shao}
\email{dingyu.shao@physics.ucla.edu}
\affiliation{Department of Physics and Astronomy, University of California, Los Angeles, California 90095, USA}      
\affiliation{Mani L. Bhaumik Institute for Theoretical Physics,
University of California, Los Angeles, California 90095, USA}            
\affiliation{Center for Frontiers in Nuclear Science, Stony Brook University, Stony Brook, New York 11794, USA}

\begin{abstract}
We propose the jet charge observable as a novel probe of flavor structure in the nucleon spin program at the Electron Ion Collider (EIC) and develop the underlying framework from first principles. We show that jet charge measurements can substantially enhance the sensitivity of spin asymmetries to different partonic flavors in the nucleon. This sensitivity can be further improved by constructing the jet charge using only a subset of hadron species (pions or kaons) in the jet. As an example, we use the Sivers asymmetry in back-to-back electron-jet production at the EIC to show that the jet charge can be a unique tool in constraining the Sivers function for different partonic flavors.
\end{abstract}

\maketitle

{\it Introduction}. The femtoscale structure of the nucleon is a central mission of current experimental programs at accelerator facilities such as the Relativistic Heavy Ion Collider (RHIC) at Brookhaven National Lab, 12 GeV CEBAF at Jefferson Lab, COMPASS at CERN, and HERMES at DESY. It is also one of the major scientific pillars of the future Electron Ion Collider (EIC)~\cite{Accardi:2012qut,Aidala:2020mzt}. In particular, the flavor and spin structure of the nucleon in terms of both one dimensional and three-dimensional (3D) imaging provides fascinating glimpses into the  nontrivial non-perturbative QCD dynamics. 

One major tool to effectively deconvolute the experimental information on the flavor and spin structure of the nucleon, as encoded in unpolarized and polarized parton distribution functions (PDFs) and their extensions such as transverse momentum distributions (TMDs), is the ``global QCD analysis''. It treats all available probes simultaneously to extract the universal PDF sets. For example, for the polarized PDFs, one relies on polarized inclusive deep inelastic scattering (DIS), semi-inclusive DIS (SIDIS) and proton-proton collisions~\cite{deFlorian:2019zkl}. In general, in fully inclusive DIS data it is difficult to disentangle different quark flavors, unless one uses different targets such as the proton, deuteron, and $^3{\rm He}$~\cite{Milner:2018mmz,Maxwell:2019htc}. In this regard, the SIDIS process is crucial: it uses the detection of hadrons in the final state as a flavor tag of the initial-state PDFs, provided the fragmentation functions are well known~\cite{Sato:2019yez}. 

In recent years, measurements at RHIC and LHC demonstrate that jets can be a useful probe for the spin structure of the nucleon~\cite{Aschenauer:2016our,Adamczyk:2017wld,Boer:2014lka}. The advent of the ElC with its high luminosity and polarized beams will unlock the full potential of jets as novel tools for probing nucleon structure. It is thus not a surprise that jet physics at the EIC has become a fast emerging research field~\cite{Liu:2018trl,Arratia:2020nxw,Liu:2020dct,Aschenauer:2019uex,Arratia:2019vju,Arratia:2020ssx,Kang:2020xyq,Aschenauer:2017jsk,Zheng:2018ssm}. While great progress has been made in jet physics at the EIC,  flavor separation in jet production and the associated spin asymmetries presents a major challenge and has not been addressed in the literature. Measurements on jets are typically inclusive over the final-state hadrons. As a consequence, although jets are better proxies of parton-level dynamics, they typically lack flavor tagging. However, it is well-known that flavor separation is essential in mapping out the flavor and spin structure of the nucleon at the EIC. Within this context, there have been some recent works on measuring hadron distributions inside jets~\cite{Kang:2020xyq,Kang:2017btw} to allow for some amount of flavor separation.
                                                                           
In this Letter, we propose the jet charge observable~\cite{Field:1977fa} as a novel flavor prism for jet observables at the EIC, especially for spin asymmetries in jet production. Jet charge has been studied extensively at the LHC~\cite{Aad:2015cua,Sirunyan:2017tyr,Li:2019dre,Chen:2019gqo,Sirunyan:2020qvi}, and similar directions at the RHIC are under exploration~\cite{Aschenauer:2015eha,bnltalk}. We develop a theoretical framework which allows us to distinguish flavor in spin asymmetries, enabling jet processes to reach their full potential in probing the flavor and spin structure of the nucleon. As concrete examples, we demonstrate for the first time the use of the jet charge observable to provide enhanced $u$- and $d$-quark flavor sensitivity for both the unpolarized quark TMDs and the polarized Sivers TMD functions using back-to-back electron-jet production at the EIC. The unpolarized TMDs are essential for the 3D imaging of the nucleon and the Sivers TMD functions, in addition, encode quantum correlations between the motion of partons and the spin of the proton. 

{\it Jet charge}. 
The jet electric charge is constructed from the definition~\cite{Krohn:2012fg, Waalewijn:2012sv}
\bea\label{eq:jetcharge}
Q_\kappa  \equiv \sum_{h\in {\rm jet}} z_h^\kappa Q_h \,,
\eea
where $\kappa$ is a parameter and $z_h = p_{T}^h/p_{T}$, where $p_{T}$ and $p_{T}^h$ denote the transverse momenta of the jet and the hadron $h$ with electric charge $Q_h$, respectively. 
The factor $z_h^\kappa$ with $\kappa>0$ suppresses the contribution of soft hadrons in the jet. Different values of $\kappa$ lead to different shapes for the jet charge distributions. It is found that the peaks in these distributions for jets initiated by different quark flavors are {better} separated for $\kappa\sim 0.3$~\cite{ATLAS-CONF-2013-086}. Correspondingly, it is the default value used in our analysis.
Note that the definition of $Q_\kappa$ allows for the option of restricting the sum over $h$ in the jet to  include only particular species, such as pions or kaons only. This flexibility can be exploited for additional flavor separation.

{\it Unpolarized TMDs}. We first consider the unpolarized electron-proton scattering process in the electron-proton center-of-mass (CM) frame, $e+p \to e + {\rm jet}+ X$, in the back-to-back region where the electron-jet transverse momentum imbalance $q_{T} = |\vec{p}_{T}^{\>e} + \vec{p}_{T}| \ll p_T^{\>e}\sim p_T$, is small. Here $p_T^{\>e}$ and $ p_T$ are the transverse momenta of the outgoing electron and the jet, respectively. The contribution to the cross section from {parton $i=u,d,\cdots$} in the proton is given by~\cite{Liu:2018trl, Arratia:2020nxw,Liu:2020dct}
\bea\label{eq:unpolarized}
\frac{\mathrm{d}^5 \sigma_{UU}^i}
{\mathrm{d}y_e \mathrm{d}^2p_{T}^e \mathrm{d}^2q_T}
=  \sigma_0 \, e_{i}^2 \,
\int \frac{\mathrm{d}^2b_T}{(2\pi)^2}
e^{iq_T\cdot b_T} \tilde{W}_{i}   \,,
\eea
where 
$y_e$ denotes the out-going lepton rapidity and 
$b_T$ is the transverse vector in the Fourier space conjugate to $q_T$. $e_i$ is the fractional charge carried by parton {$i$} and $\sigma_0$~\cite{Arratia:2020nxw} is the Born cross section for the partonic subprocess.
The spin-averaged structure function $\tilde{W}_{i}$ for parton $i$ has the factorized form
\bea
\label{eq:Wi}
  \tilde{W}_{i}
 = \tilde{f}_{i} (x,b_T, \mu)
 S_J(b_T,R, \mu)
 H(Q,\mu)
 {\cal J}_{i}(p_T R, \mu)  \,,
\eea
where $\tilde{f}_{i} (x,b_T, \mu)$ is the unpolarized TMD function~\cite{Collins:2011zzd}, $H$ is the hard function, $S_J$ is the soft factor~\cite{Liu:2018trl, Arratia:2020nxw,Liu:2020dct}, and $Q$ is the invariant mass of the virtual photon. ${\cal J}_{i}(p_T R, \mu)$ is the universal jet function that describes the dynamics of the jet with radius $R$ initiated by the struck parton of flavor $i$~\cite{Sterman:1977wj,Jager:2004jh,Ellis:2010rwa}. Eq.~\eqref{eq:Wi} involves the factorization scale $\mu$, and also implicitly depends on another scale $\mu_b$~\cite{Arratia:2020nxw}, with central values $p_T$ and $2e^{-\gamma_E}/b_T$, respectively, where $\gamma_E$ is the Euler constant.

The measured cross section includes a sum over all partonic channels so that $\mathrm{d}\sigma_{UU}=\sum_{i} \mathrm{d}\sigma_{UU}^i$. Thus, it is sensitive only to a specific flavor combination of TMDs. However, by making an additional  measurement of the jet charge and  grouping the data into jet charge bins,  one can enhance or suppress sensitivity to TMDs of different flavors. The generalization of Eq.~(\ref{eq:unpolarized}) in which the jet charge $Q_\kappa$ is also measured is achieved by replacing ${\cal J}_{i}(p_T R, \mu)$ in $\tilde{W}_i$ in Eq.~(\ref{eq:Wi}) with the function ${\cal G}_{i}(Q_\kappa, p_T R, \mu)$~\cite{Waalewijn:2012sv}, leading to the result:
\bea
\label{eq:jetchargefunctionnorm}
\frac{\mathrm{d}^6 \sigma_{UU}^i}{\mathrm{d}y_e\mathrm{d}^2p_{T}^e\mathrm{d}^2q_T dQ_\kappa}
= \frac{\mathrm{d}^5 \sigma_{UU}^i}{\mathrm{d}y_e\mathrm{d}^2p_{T}^e\mathrm{d}^2q_T} \, \frac{{\cal G}_{i}(Q_\kappa, p_T R, \mu)}{{\cal J}_{i}(p_T R, \mu)}\,.
\eea
Note that $\int \mathrm{d}Q_\kappa \>{\cal G}_{i}(Q_\kappa,p_T R, \mu)={\cal J}_{i}(p_T R, \mu)$,
so that integrating over all possible jet charge values gives back the standard jet cross section.

Since Eq.~\eqref{eq:jetchargefunctionnorm} is achieved simply by the replacement ${\cal J}_i\to {\cal G}_i$, the renormalization scale independence of the cross section implies that the renormalization group evolution properties of ${\cal G}_i$ are exactly the same as ${\cal J}_i$~\cite{Liu:2012sz,Arratia:2020nxw}.
The universality of both jet functions ${\cal J}_i$ and ${\cal G}_i$ implies that they also appear in the factorization formulae of different jet processes in $pp$~\cite{Liu:2012sz,Sun:2018icb,Buffing:2018ggv,Kang:2019ahe,Chien:2019gyf,Liu:2017pbb} and $e^+e^-$ collisions~\cite{Ellis:2010rwa}. 

For jet charge binned data, it’s useful to introduce the $N$-th moment for a jet initiated by the parton $i$
\bea
\label{eq:QNbin}
\langle Q_\kappa^N \rangle_{i,{\rm bin}}  =
\int_{Q_\kappa\in {\rm bin}}  \mathrm{d}Q_\kappa    
\, Q_\kappa^N \, 
 \frac{{\cal G}_{i}(Q_\kappa, p_T R,  \mu)}{{\cal J}_i(p_T R, \mu)}  \,.
\eea
In the rest of the Letter, we are particularly interested in the $0$-th moment $r_{i,{\rm bin}} \equiv 
\langle Q_\kappa^0 \rangle_{i,{\rm bin}} $
which, 
as seen from Eqs.~\eqref{eq:jetchargefunctionnorm} and~\eqref{eq:QNbin}, gives the fractional contribution in a given jet charge bin and thus satisfies the sum rule $\sum_{\rm bins} r_{i,\rm bin} = 1$. 
The differential cross section restricted to the particular jet charge bin is then given by
\bea\label{eq:chargebin}
\frac{\mathrm{d}^5 \sigma_{UU,\,\rm bin}}
{\mathrm{d}y_e \mathrm{d}^2p_{T}^e \mathrm{d}^2q_T}
=  
\sum_{i=u,d,\cdots}\, r_{i,\rm bin} \,  
\frac{\mathrm{d}^5 \sigma_{UU}^i }
{\mathrm{d}y_e \mathrm{d}^2p_{T}^e \mathrm{d}^2q_T} \,,
\eea
which follows from a combination of Eqs.~\eqref{eq:jetchargefunctionnorm} and~\eqref{eq:QNbin} with $N=0$. This is a key result that allows for flavor separation of the TMDs through an appropriate selection of jet charge bins that can enhance the contribution of the $i$-th parton flavor depending on the value of $r_{i,\rm bin}$.

Note that the universality of ${\cal G}_i$ and ${\cal J}_i$ make the charge bin fraction $r_{i,\rm bin}$ process independent, 
allowing for their extraction from a global analysis that makes use of binned jet charge measurements  in  the electron-proton scattering, $pp$ collisions, and $e^+e^-$ annihilation processes with appropriate kinematic cuts.  Furthermore, the $r_{i,\rm bin}$ are $\mu$-independent due to a cancellation of the scale dependence between ${\cal G}_{i}$ and ${\cal J}_i$ in Eq.~(\ref{eq:QNbin}).

The $r_{i,{\rm bin}}$ does however have a mild dependence on $p_{T} R$ via loop suppressed effects~\cite{Waalewijn:2012sv}.  This loop suppression can be understood by noting that at tree-level the jet consists of single parton evolving into hadrons. The $p_TR$ dependence first arises at $\alpha_s$ through a perturbative splitting within the jet before hadronization. We have checked this behavior via Pythia8 \cite{Sjostrand:2007gs} simulations over a wide range of $p_{T}R$ for $\kappa = 0.3$. This property can allow for the extraction of the $r_{i,{\rm bin}}$ from jet observables over a wide range of kinematics. In addition, the sum rule for $r_{i,{\rm bin}}$ provides an additional constraint.

The $r_{i,{\rm bin}}$ of different species $i$ will also be related by QCD flavor and charge conjugation symmetries,  further reducing the number of fit parameters.
Furthermore, as seen in Eq.~(\ref{eq:chargebin}), the $r_{i,\rm bin}$ only affect the relative size  of each partonic contribution and not the shapes of their kinematic distributions. 

All these features facilitate the extraction of the $r_{i,\rm bin}$ through a global analysis. In fact, measurements of the jet charge distribution have been carried out in both $pp$ and $Pb Pb$ collisions at the LHC~\cite{Aad:2015cua,Sirunyan:2017tyr,Sirunyan:2020qvi}.

{\it Single Spin Asymmetry}.  
One can generalize the factorization to the process 
where the proton is transversely polarized with spin vector $\vec{S}_\perp$. In this case, $\mathrm{d}\sigma(S_\perp)=\mathrm{d}\sigma_{UU}+\mathrm{d}\sigma_{UT}(S_\perp)$, where the spin-dependent $\mathrm{d}\sigma_{UT}$ depends on the Sivers function~\cite{Sivers:1989cc}. 
Following the same steps used to  arrive  at Eq.~(\ref{eq:chargebin}) for the unpolarized case, we find that the transverse-spin dependent cross section within a jet charge bin can be related to the standard transverse-spin dependent cross section by
\bea\label{eq:polarizedcharge}
\hspace{-2.2ex} \frac{\mathrm{d}^5 \sigma_{UT,{\rm bin}}(S_\perp)  }
{\mathrm{d}y_e \mathrm{d}^2p^e_{T} \mathrm{d}^2q_T} 
= 
\sum_{i=u,d,\cdots} 
r_{i,\rm bin} 
\frac{\mathrm{d}^5 \sigma^i_{UT}  ( S_\perp) }
{\mathrm{d}y_e \mathrm{d}^2p^e_{T} \mathrm{d}^2q_T}\,, 
\eea
where $r_{i,\rm bin}$ is the same 
as in the unpolarized case. 
The standard transverse-spin dependent cross section for a given partonic channel $i$ reads~\cite{Liu:2018trl, Arratia:2020nxw,Liu:2020dct}

\bea
\hspace{-1.ex}
\frac{\mathrm{d}^5\sigma_{UT}^i  ( S_\perp) }
{\mathrm{d}y_e \mathrm{d}^2p_{T}^e \mathrm{d}^2q_T}
=  e_{i}^2 \, \sigma_0 \, \epsilon_{\alpha\beta}S_{\perp}^{\alpha}
\int \frac{\mathrm{d}^2b_T}{(2\pi)^2}
e^{iq_T\cdot b_T} \tilde{W}^{\beta}_{T,i} \,,
\eea
where the spin-dependent structure function $\tilde{W}^\beta_{T,i}$ takes the factorized form
\bea
  \tilde{W}^{\beta}_{T,i}
=  \tilde{f}_{1T,i}^{\perp,\beta} (x,b_T,\mu)
 S_J(b_T,R, \mu)
 H(Q,\mu) {\cal J}_i(p_T R,\mu) \,,
 \nonumber
\eea
where $ \tilde{f}_{1T,i}^{\perp,\beta}(x,b_T, \mu)$ is the Sivers function~\cite{Liu:2018trl, Arratia:2020nxw,Liu:2020dct}, which can be directly accessed via the single spin asymmetry 
\bea
A_{UT} = \frac{\mathrm{d}\sigma(S_\perp^\uparrow) -\mathrm{d} \sigma(S_\perp^\downarrow)}{\mathrm{d} \sigma(S_\perp^\uparrow) + \mathrm{d} \sigma(S_\perp^\downarrow)} = \frac{\mathrm{d}\sigma_{UT}}{\mathrm{d} \sigma_{UU}}.
\eea
We denote the spin asymmetry in a particular jet charge bin as $A_{UT}^{\rm bin}$.

Similar to Eq.~(\ref{eq:chargebin}), Eq.~(\ref{eq:polarizedcharge}) is another key result that allows for flavor separation of the Sivers function. The dominant contribution to $A_{UT}$ is known to come from the $u$-quark channel~\cite{Arratia:2020nxw,Bacchetta:2006tn}, 
resulting in sensitivity primarily to the $u$-quark Sivers function. However, following Eq.~(\ref{eq:polarizedcharge}), an appropriate selection of jet charge bins could 
enhance the sensitivity of $A^{\rm bin}_{UT}$ to the Sivers functions of the other quark flavors.

\begin{figure}[b!]
\vspace{-5.ex}
  \centering
\includegraphics[width=.95\linewidth]{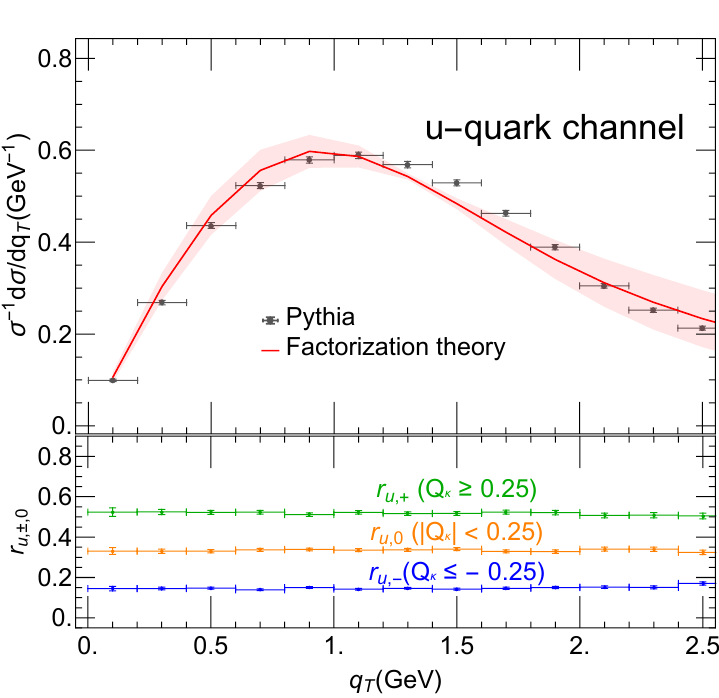} 
 \vspace{-2.ex}
\caption{The contribution of the $u$-quark TMD to the normalized $q_T$-distribution $\sigma^{-1}\mathrm{d}\sigma/\mathrm{d}q_T$. The band shows the scale uncertainty.}
 \vspace{-2.ex}
\label{fig:u-jet}
\end{figure}

 {\it Phenomenology.} We present numerical results to demonstrate the implications of the key results in Eqs.~(\ref{eq:chargebin}) and (\ref{eq:polarizedcharge}). We work 
 with CM energy $\sqrt{s} = 105 \> {\rm GeV}$ and apply event selection cuts of $Q^2 \ge 10 \> {\rm GeV}^2$,  $0.1\le y\le 0.9$, $15 \> {\rm GeV} \le p_{T}^e  \le 20 \>{\rm GeV}$,   $q_T \le 2.5\>{\rm GeV}$, and $10 \> {\rm GeV} \le p_{T}  \le 25 \>{\rm GeV}$, where $y$ denotes the inelasticity. Jets are constructed using anti-$k_T$ jet algorithm~\cite{Cacciari:2011ma} with radius $R = 1$. For the back-to-back electron-jet production, the gluon TMD does not contribute so that the quark TMDs can be probed cleanly. In the rest of the paper, we present the normalized $q_T$-distribution, $\sigma^{-1}\mathrm{d}\sigma/\mathrm{d}q_T$, with $y_e$ and $p_T^e$ integrated over the allowed range corresponding to the above selection cuts. We include theoretical uncertainties by varying $\mu$ and $\mu_b$ by a factor of 2 around their central values and taking the envelop.
 
 
We first present the unpolarized studies. For calibration purposes, in the upper panel of Fig.~\ref{fig:u-jet} we check the consistency between Pythia8 simulations and the theoretical predictions, only including the $u$-quark contribution. Eq.~(\ref{eq:unpolarized}) is evaluated at the next-to-leading logarithmic accuracy, 
with non-global logarithms included~\cite{Liu:2018trl, Arratia:2020nxw,Liu:2020dct,Becher:2015hka,Chien:2019gyf}. We parameterize the $u$-quark TMD following Ref.~\cite{Arratia:2020nxw}.  In Fig.~\ref{fig:u-jet}, we see good agreement between simulation and theory for the normalized $q_T$-distribution.  

Next, we study the $q_T$-distribution in different jet charge bins. The jet charges are constructed using Eq.~(\ref{eq:jetcharge}) with $\kappa=0.3$ and only including charged pions in the sum over hadrons. Thus, if a jet contains no charged pions, its charge  vanishes. We divide the data into three jet charge bins, the negative (``$-$") bin with $Q_\kappa \le- 0.25$, the neutral bin (``$0$") with $|Q_\kappa| < 0.25$, and the positive bin (``$+$") with $Q_\kappa \ge 0.25$. This jet charge bin choice is motivated by our finding that the $d$-quark jet charge distribution peaks around $Q_\kappa \sim -0.25$. See similar observation in Refs.~\cite{Waalewijn:2012sv,Krohn:2012fg}. 

The lower panel of Fig.~\ref{fig:u-jet} shows the jet charge bin fractions $r_{u,\pm,0}$ for the $u$-quark jets as a function of $q_T$. These bin fractions were determined from Pythia and are found to be independent of $q_T$. This agrees with the universality as expected from our factorization theorem.
In Tab.~\ref{tab:charge-bin-fraction}, we summarize the $r_{i,\pm,0}$ values, including other quark flavors.
\begin{table}[t]
\centering
\begin{tabular}{c | c  c c c c c }
\hline
\hline
  & $u$ & $ {\bar u}$ & $d$ & ${\bar d}$ & $s$ & ${\bar s} $  \  \\ 
\hline
 $r_{i,+}$ & $0.52$ & $ 0.17 $ & $0.15$ & $0.53$ & $0.30$ & $ 0.34 $  \  \\ 
 $r_{i,-}$ & $ 0.15$ & $ 0.49 $ & $0.52$ & $0.15$ & $0.36$ & $0.32 $  \  \\ 
 $r_{i,0}$ & $0.33$ & $ 0.35 $ & $0.33$ & $0.32$ & $0.35$ & $0.34 $  \  \\ 
\hline
\end{tabular}
\caption{The jet charge bin fractions $r_{i,\pm,0}$ for various quark flavors, obtained from Pythia8. }
\label{tab:charge-bin-fraction}
\vspace{-4.ex}
\end{table}
 We find that $r_{u,{\pm,0}} \approx r_{{\bar d},{\pm,0}} \approx r_{d,{\mp,0}}  \approx r_{{{\bar u}},{\mp,0}}$ and $r_{s,{\pm,0}} \approx r_{{{\bar s}},{\mp,0}}$, as expected from the QCD flavor and charge conjugation symmetries and the fact that only charged pions were included in constructing the jet charge.  A jet charge based only on charged kaons can increase the negative charge bin fractions of the $s$-quarks by a factor of  ${\cal O}(5)$ with respect to the $u (d)$-quark fraction, allowing for better sensitivity to the strange quark distributions.
 \begin{figure}[h!]
\vspace{-3.ex}
  \centering
\includegraphics[width=.95\linewidth]{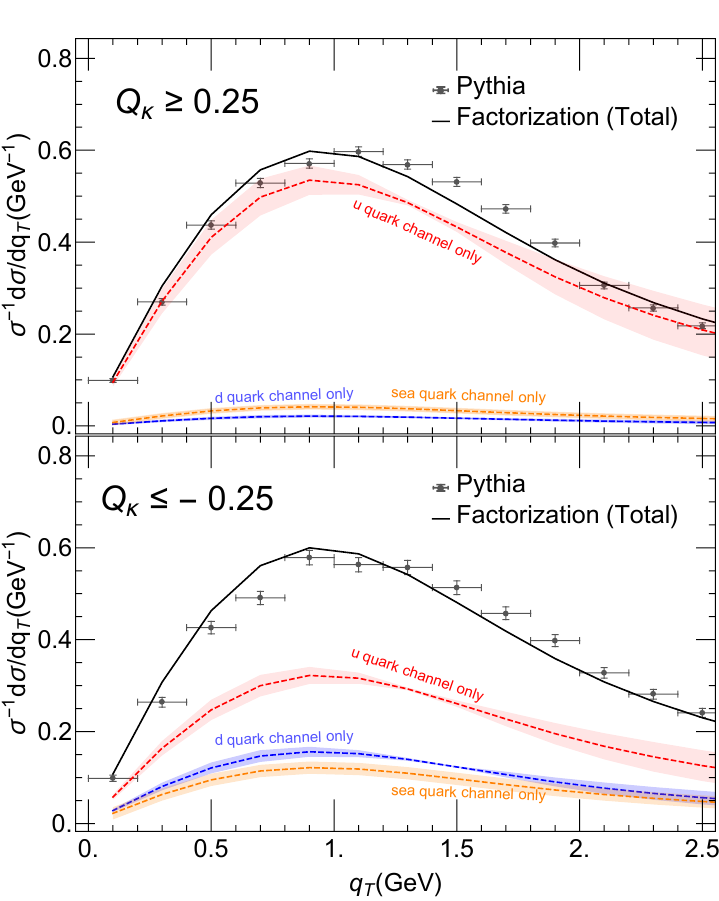} 
\vspace{-2.ex}
\caption{The relative size of contributions from the unpolarized $u$-, $d$-, and sea quark TMDs.}
 \vspace{-1.ex}
\label{fig:UPfraction}
\end{figure}
 
The values in Tab.~\ref{tab:charge-bin-fraction} are used as  inputs in the rest of the analysis to make 
predictions for the relative size of contributions from different quark flavors in each bin. In practice, the bin fractions $r_{i,{\pm,0}}$ for each parton flavor could be obtained with a fit of the the cross section in Eq.~\eqref{eq:chargebin} to the  $q_T$-spectrum in each bin, and used as inputs for 
predictions of $A_{UT}^{\rm bin}$.

From 
Tab.~\ref{tab:charge-bin-fraction}, selecting the negative bin will significantly reduce the size of the $u$-quark contribution and enhance the $d$-quark contribution, compared to the neutral and positive bins. 
This can be seen in Fig.~\ref{fig:UPfraction}, where for $Q_\kappa \geq 0.25$ (upper panel), the $d$-quark and the sea quark contributions are small and the $u$-quark contribution dominates the bulk of the distribution. While, for $Q_\kappa \leq -0.25$ (bottom panel) the $u$-quark contribution is now  suppressed and the rest species are relatively enhanced and more readily accessible in experiment. 

\begin{figure}[b!]
\vspace{-5.ex}
  \centering
\includegraphics[width=.9\linewidth]{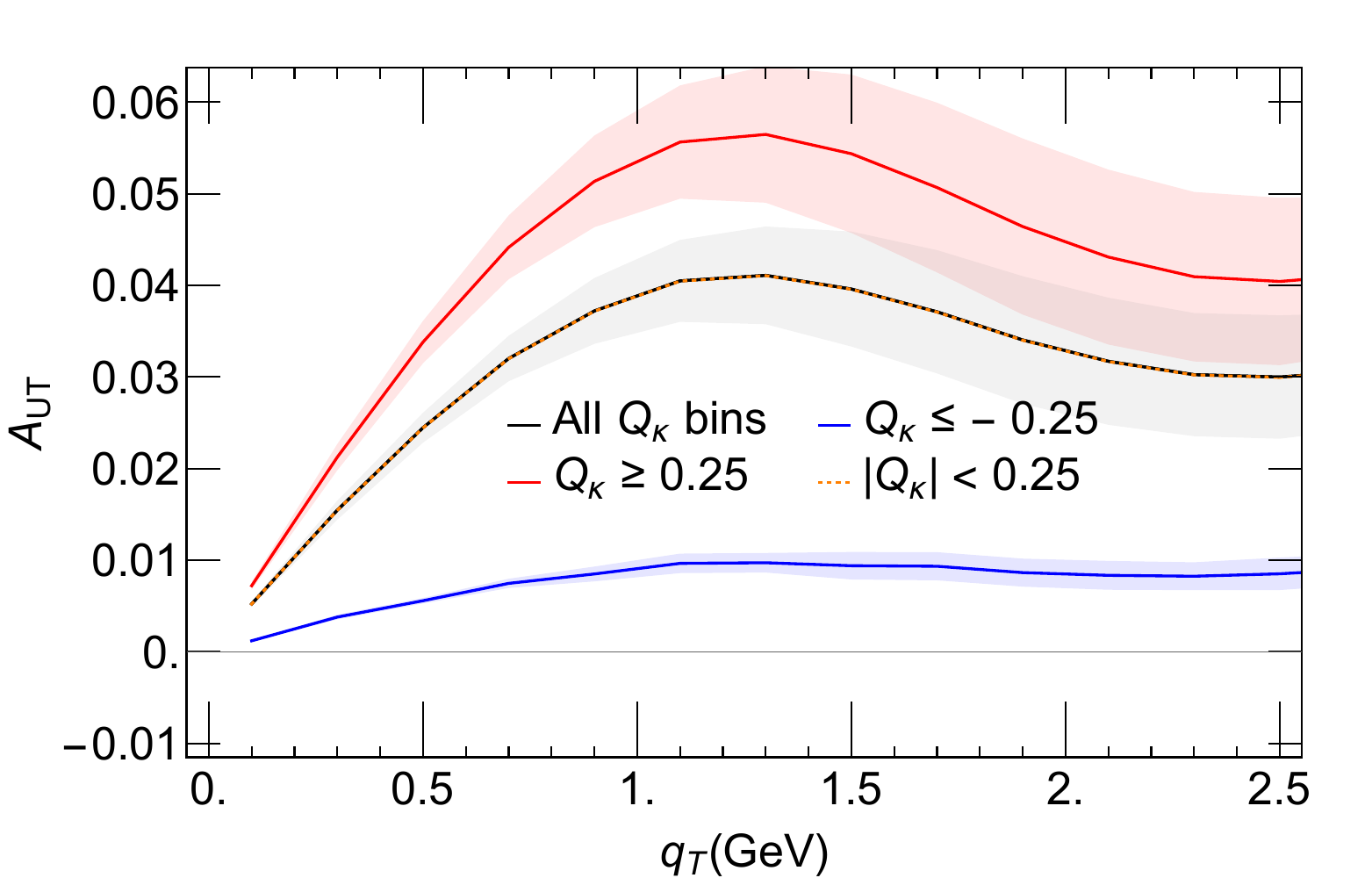} 
 \vspace{-2.ex}
\caption{
Predictions for the Sivers asymmetry in different jet charge bins.}
 \vspace{-5.ex}
\label{fig:spin-symmetry}
\end{figure}
Next, we investigate the single spin asymmetry.  Fig.~\ref{fig:spin-symmetry} shows a comparison of the theoretical predictions for the standard asymmetry $A_{UT}$ with 
$A_{UT}^{\rm bin}$. The predictions can be compared with the future measurements. 
The Sivers functions are parameterized following Ref.~\cite{Echevarria:2014xaa}, except that we ignore the strange quark Sivers functions in light of recent global analysis which shows the very small size of their contributions~\cite{Echevarria:2020hpy,Bacchetta:2020gko,Cammarota:2020qcw}. 

From Fig.~\ref{fig:spin-symmetry}, we see that $A_{UT}$ is positive and large for $Q_\kappa \ge 0.25$ which is due to the dominant and positive $u$-quark Sivers function, while the other channels are highly suppressed. In the $Q_\kappa \le - 0.25$ bin, the $u$-quark contribution is substantially reduced and comparable to the size of the $d$-quark contribution. Thus, the cancellation between the $u$- and $d$-quark Sivers functions leads to the a small spin asymmetry for $Q_\kappa \le - 0.25$. For $|Q_\kappa| < 0.25$, since the bin fractions are roughly the same for all partonic channels, as seen in Tab.~\ref{tab:charge-bin-fraction}, the spin asymmetry is close to the standard $A_{UT}$ as expected. 
Similar behavior is observed 
in SIDIS when charged pions are measured~\cite{Airapetian:2020zzo}.

\begin{figure}[htp]
\vspace{-2.ex}
  \centering
\includegraphics[width=.95\linewidth]{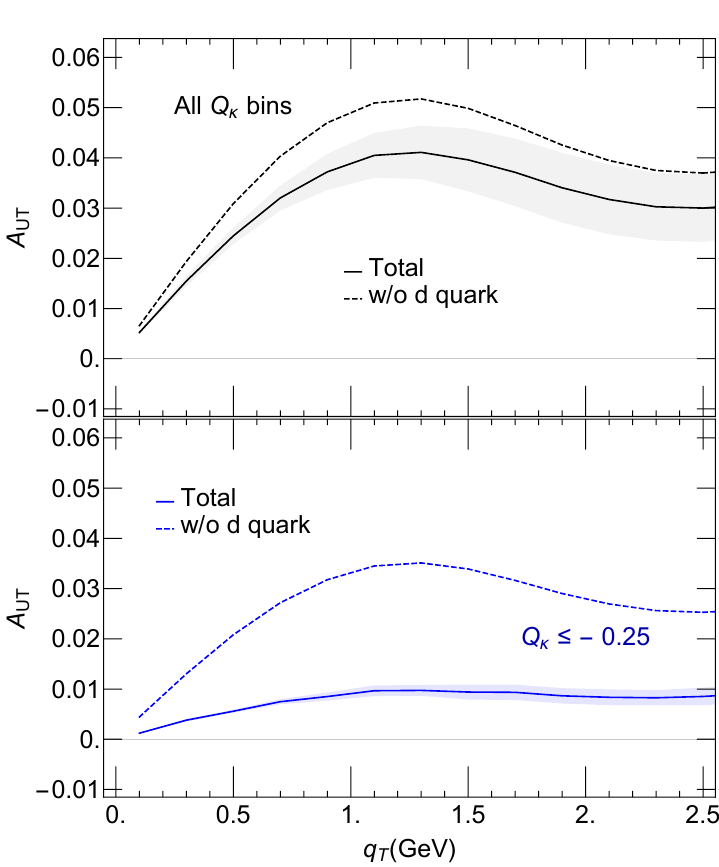} 
 \vspace{-2.ex}
\caption{Sensitivities of the 
$d$-quark channels to the Sivers asymmetry.}
\label{fig:sensitivity}
\vspace{-5.ex}
\end{figure}
Finally, we study the sensitivity of the asymmetry to different quark flavors. As seen in the top panel of Fig.~\ref{fig:sensitivity}, the standard Sivers asymmetry is not sensitive to the $d$-quark Sivers function, since the size of the asymmetry $A_{UT}$  changes only slightly when one removes the $d$-quark contribution (``w/o $d$-quark''). However, if one restricts to the $Q_\kappa \leq -0.25$ bin, removing the $d$-quark contribution leads to a significant change in the asymmetry as seen in the bottom panel of Fig.~\ref{fig:sensitivity}. This demonstrates the dramatically enhanced sensitivity to the  valence $d$-quark contribution in $A_{UT}^{\rm bin}$. 

{\it Conclusions.}  
In this Letter, we propose the jet charge as a unique flavor probe of polarized and unpolarized TMDs at the EIC. As concrete examples, we study back-to-back electron-jet production and we give predictions for the small $q_T$-distribution and the Sivers asymmetry in different jet charge bins, based on a factorization framework. In order to demonstrate its power for flavor separation, we compare the flavor sensitivities of the jet charge binned unpolarized cross section and the Sivers spin asymmetries with their standard counterparts at the EIC. We show that through an appropriate selection of the charge bins, the sensitivity to the unpolarized TMDs and Sivers functions of different quark species can be enhanced. The flavor sensitivity can be further improved when using specific subset of hadron species to define the jet charge. 
Other possible applications of the jet charge include probing the spin-dependent collinear PDFs, like the helicity distribution~\cite{Boughezal:2018azh,Borsa:2020ulb} and the Qiu-Sterman function~\cite{Qiu:1991pp,Aschenauer:2015eha}.
We expect the ideas proposed in this work to open new directions of exploration for jet/spin physics at the future EIC and complement other probes of the flavor information via the conventional SIDIS process.

\begin{acknowledgments}
{\it Acknowledgements.} We thank Nobuo Sato for valuable discussions. This work is supported by the National Science Foundation under Contract No.~PHY-1720486 and CAREER award~PHY-1945471 (Z.K., D.Y.S.), by the National Natural Science Foundation of China under Grant No.~11775023 (X.L.), and by Center for Frontiers in Nuclear Science of Stony Brook University and Brookhaven National Laboratory (D.Y.S.).
\end{acknowledgments}

\bibliography{references}

\bibliographystyle{h-physrev5}

\end{document}